 \documentclass[12pt]{iopart}

\usepackage{color}
\newcommand{\comma}{\;\; ,}
\newcommand{\period}{\;\; .}
\newcommand{\eq}{\; = \;}
\newcommand{\sep}{\;\; , \;\;}

\newcommand{\be}{\begin{equation}}
\newcommand{\bd}{\begin{displaymath}}
\newcommand{\ee}{\end{equation}}
\newcommand{\ed}{\end{displaymath}}
\newcommand{\ba}{\begin{eqnarray}}
\newcommand{\ea}{\end{eqnarray}}

\newcommand{\Wb}{\overline{W}}
\newcommand{\Kb}{\overline{K}}

\renewcommand{\i}{{\rm i}}
\newcommand{\ex}{{\rm e}}

\newcommand{\ctnclr}{blue}
\newcommand{\om}{\omega}

\begin{document}

 \title[Developments in exact solutions in 
statistical mechanics ]{Some comments on developments in exact solutions in 
statistical mechanics since 1944}


\author{ R.J. Baxter\\
{\protect \small  Mathematical
Sciences Institute}\\
{\protect \small  The Australian National University,
 Canberra, A.C.T. 0200, Australia}}

\date{}

\definecolor{magenta}{rgb}{0.5,0,0.5}

\definecolor{red}{rgb}{0.8,0,0}

\definecolor{green}{rgb}{0,0.5,0}

\definecolor{blue}{rgb}{0,0,0.8}

\definecolor{black}{rgb}{0.4,0.4,0.4}

\definecolor{brown}{rgb}{0.4,0.2,0.2}

 \setlength{\unitlength}{1pt}


 \begin{abstract}
 Lars Onsager and Bruria Kaufman calculated the partition
  function of the  Ising model exactly in 1944 and 1949.  Since then
there have been  many developments in the exact solution of 
similar, but usually  more complicated, models. Here I shall mention 
a few, and show how some of the latest work seems to be returning
once again to the properties observed by Onsager and Kaufman.
\end{abstract}

 
  {{\bf ArXiv ePrint: } \color{blue} 1010.0710}



{{\color{\ctnclr}

 \tableofcontents 
 \title[Developments in exact solutions in statistical mechanics ]{}
 }

{{\color{\ctnclr}

  \section{Introduction}}
  

\setcounter{equation}{0}


The first solution of a finite-dimensional lattice model that exhibits a 
phase  transition was the calculation by Lars Onsager in 1944 of the 
partition function  of the zero-field square-lattice Ising 
model.{\color{\ctnclr}\cite{Onsager1944}} In 1949 Bruria Kaufman
derived this result using spinor or free-fermion 
operators.{\color{\ctnclr}\cite{Kaufman1949}}


We discuss the history of the derivation of the spontaneous 
magnetization ${\cal M}$ of the Ising model in section 
 {\color{blue} 4}: it depended
on first using the free fermion structure to write $\cal M$, for a 
finite lattice, in terms of  a determinant, then
taking the limit when the lattice and the dimensions of
the determinant become 
infinite.{\color{\ctnclr}\cite{Onsager1949, Yang1952,MPW1963}}

After the Ising model, and the related dimer problems,  the next 
models to be solved were the six-vertex and  eight-vertex 
models. We discuss these in sections {\color{blue} 5} and 
 {\color{blue} 6}, in particular we outline
the Bethe ansatz calculation for the general six-vertex model 
in a field.Then many other solvable models were found, both
two  and
three-dimensional.{\color{\ctnclr}\cite{Lieb1}-\cite{BazBax1993}}
These were more complicated than the Ising model, lacking 
its simple  free-fermion structure and explicit determinantal 
expressions for correlations such as the magnetization.
(One exception is the six-vertex model with particular boundary 
conditions.{\color{\ctnclr}\cite{Korepin1982}})
 
 One of the most challenging has been the two-dimensional  
 $N$-state solvable chiral Potts  model.{\color{\ctnclr} \cite{BPY1988}}
  It  lacks the ``rapidity-difference''  property, so there is
 no explicit single-valued parametrization of the Boltzmann 
 weights of   the model, and the simple inversion relation and 
 corner transfer matrix  tricks to calculate the free energy and 
 magnetization fail.
 

 However, there is a special  ``superintegrable''  
 case{\color{\ctnclr} \cite{RJB1988}}  of the chiral
 Potts model whose magnetization 
 is that of the general model  and which has simple properties similar
  to those of the Ising model.  Indeed, when $N = 2$ it  {\em is} the 
  Ising model.
  
 
 In particular it has recently been 
 shown{\color{\ctnclr} \cite{paper1}-\cite{paper5}}
 that its magnetization can be expressed as an $m$ by $m$  
 determinant $D$, even for a  finite lattice.

So the wheel has come full circle and we are back to
determinantal expressions. It still remains to calculate
 $D$ in the large-lattice limit, when $m$ becomes 
infinite. For the superintegrable chiral Potts model, the 
determinant is not Toeplitz, but it is Cauchy-like, so $D$ can 
be  evaluated explicitly and the magnetization 
obtained.{\color{\ctnclr} \cite{paper4}} 


{{\color{\ctnclr}
\vspace{3mm}

\section{Partition functions and transfer matrices}
 \setcounter{equation}{0}}
 
 We draw the square lattice diagonally, as in Figure 
 {\color{\ctnclr}  \ref{sqlatt}}, with 
 $M$ rows of sites  and $L$ sites per row. On each site $i$
 we place a ``spin'' $\sigma_i$, which takes some set of discrete 
 values.  For the Ising model, $\sigma_i = +1 $ or $-1$; for a general  
 $N$-state  model, $\sigma_i = 0, 1, \ldots, N-1$. Adjacent spins 
 $i, j$ interact with an  energy $\epsilon (\sigma_i, \sigma_j)$. 
 Spins on the top and bottom   rows are fixed to the value $0$
(for the Ising model, to the value $+1$), and  cylindrical boundary 
conditions are imposed, so that we identify the 
spins in the extreme left column with   those in the extreme right.


 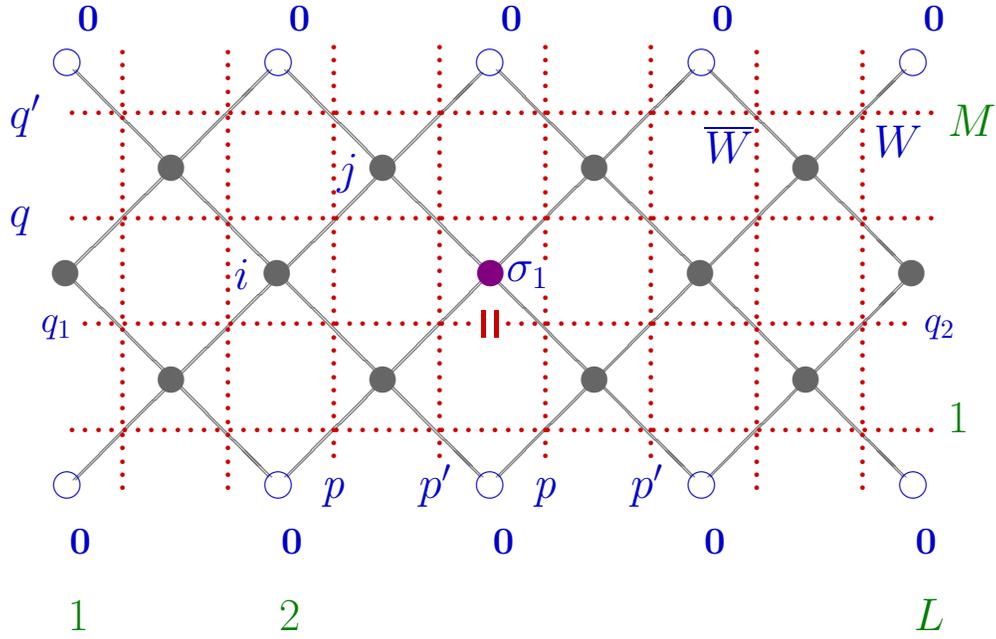
\begin{figure}[hbt]
\begin{picture}(320,270) (-34,-40)

\setlength{\unitlength}{1.0pt}

{\color{black}

\put (21,100) {\line(1,1) {77}}
\put (22,100) {\line(1,1) {77}}

\put (24,22) {\line(1,1) {154}}
\put (25,22) {\line(1,1) {154}}

\put (104,23) {\line(1,1) {154}}
\put (105,23) {\line(1,1) {154}}

\put (184,23) {\line(1,1) {154}}
\put (185,23) {\line(1,1) {154}}

\put (264,23) {\line(1,1) {75}}
\put (265,23) {\line(1,1) {75}}

\put (24,97) {\line(1,-1) {74}}
\put (25,97) {\line(1,-1) {74}}

\put (24,177) {\line(1,-1) {154}}
\put (25,177) {\line(1,-1) {154}}

\put (104,177) {\line(1,-1) {154}}
\put (105,177) {\line(1,-1) {154}}

\put (184,177) {\line(1,-1) {154}}
\put (185,177) {\line(1,-1) {154}}

\put (264,177) {\line(1,-1) {74}}
\put (265,177) {\line(1,-1) {74}}

\multiput(21,100)(80,0){2}{\circle*{10}}
\multiput(261,100)(80,0){2}{\circle*{10}}
\multiput(61,140)(80,0){4}{\circle*{10}}

\multiput(61,60)(80,0){4}{\circle*{10}}
}

{\color{green}
\put (351,152){\Large  {$ M $}}
\put (351,40){\Large  $  1$}
\put (18,-35){\Large   $1$ }
\put (98,-35){\Large   $2$ }
\put (338,-35){\Large   $L$ }

}

{\color{blue}

\put (116,134) {\Large $j$}
\put (77,94) {\Large $i$}
\put (180,97) {\Large $\sigma_1$}

\put (227,15){\Large {$ p'$}}
\put (190,15){\Large {\it p}}
\put (147,15){\Large {$ p'$}}
\put (110,15){\Large {\it p}}
\put (-8,118){\Large {$ q $}}
\put (-8,155){\Large {$ q' $}}

\put (4,78){\large {$ q_1 $}}
\put (338,78){\large {$ q_2 $}}

\put(255,142) {\Large $\Wb$}
\put(319,144) {\Large $ W$}

}

{\color{magenta}
\put (170,100) {\circle*{10}}

}

{\color{red}

\multiput(6,160)(5,0){66}{\bf .}
\multiput(6,120)(5,0){66}{\bf .}
\multiput(11,80)(5,0){30}{\bf .}
\multiput(171,80)(5,0){31}{\bf .}
\multiput(6,40)(5,0){66}{\bf .}

\multiput(25,18)(0,5){34}{\bf .}
\multiput(65,18)(0,5){34}{\bf .}
\multiput(105,30)(0,5){32}{\bf .}
\multiput(145,30)(0,5){32}{\bf .}
\multiput(185,30)(0,5){13}{\bf .}
\multiput(185,105)(0,5){17}{\bf .}
\multiput(225,30)(0,5){32}{\bf .}
\multiput(265,18)(0,5){34}{\bf .}
\multiput(305,18)(0,5){34}{\bf .}

\put(163,76)  {\line(0,1) {10}}
\put(164,76)  {\line(0,1) {10}}

\put(168,76)  {\line(0,1) {10}}
\put(169,76)  {\line(0,1) {10}}

}

{\color{blue}
\put (6,192){\large  \bf 0}
\put (86,192){\large  \bf  0}
\put (166,192){\large  \bf 0}
\put (246,192){\large  \bf  0}
\put (326,192){\large  \bf  0}

\put (3,-6){\large \bf  0}
\put (83,-6){\large \bf  0}
\put (163,-6){\large  \bf  0}
\put (243,-6){\large  \bf  0}
\put (323,-6){\large  \bf  0}

\multiput(2,20)(80,0){5}{\circle{10}}
\multiput(2,180)(80,0){5}{\circle{10}}
}
\end{picture}

 \caption{\small The square lattice turned 
 through $45^{\circ}$.} 
  \label{sqlatt}
  \vspace{1cm}

 \end{figure}



  Let
  \bd W(\sigma_i, \sigma_j) \eq 
  \ex^{-\epsilon (\sigma_i, \sigma_j)/k_B {\cal T} } \ed
  be the Boltzmann weight of the edge $\langle i,j \rangle$ 
  ($k_B$ is Boltzmann's   constant and $\cal T$ the temperature). 
  Then the partition function  is
  \be \label{partfn}
  Z   \eq  \sum_{\sigma} \, \prod_{\langle  i,j \rangle} 
  W(\sigma_i, \sigma_j)    \comma \ee
  where  $\sigma$ is the set of all spins, and the sum is over all their 
  permitted values.
  
  We are particularly interested in calculating the partition function 
  per site
  \be \kappa \eq Z^{1/LM} \comma  \ee
  the dimensionless free energy $f = - \log \kappa$,
  and averages such as the Ising model magnetization
  \be \label{defIsingmag}
  {\cal M} \eq Z^{-1}  \sum_{\sigma} \,\sigma_1 \, 
   \prod_{\langle  i,j \rangle} W(\sigma_i, \sigma_j)  \period  \ee
  We expect $\kappa$ and $\cal M$ to tend to limits when 
  $L, M \rightarrow \infty$ and the central spin $\sigma_1$ 
  becomes  deep within the lattice.
  

  Of course, such calculations are not easy: for the Ising model on a 
  ten by   ten lattice there are $2^{100} \sim 10^{30}$ terms in the sum 
  in {\color{\ctnclr} (\ref{partfn})}.  One starts by defining the row-to-row  transfer matrix 
  $T$. Let  $s = \{ \sigma_1, \sigma_2, \ldots, \sigma_L \}$ be the spins 
  in one row  of the lattice, and   
  $s' = \{ \sigma'_1, \sigma'_2, \ldots, \sigma'_L \}$
  the spins in the row above. Define the $N^L$ by $N^L$ matrix $T$
  with entries
  \be T_{s,s'} \eq \prod_i \,  W(\sigma_i, \sigma'_i ) \, {\overline W}
   (\sigma_i, \sigma'_{i-1} ) \comma \ee
   taking $W$ to be the weight on SW-NE edges and $\overline{W}$ 
   to be  the (possibly different) weight on SE-NW edges. Then the
   partition  function is
   \be \label{partfn2}
   Z \eq u^{\dagger} T^M  u  \comma \ee
   where $u$ is the $N^L$-dimensional vectors with entries
   \be  \label{defu}
   u_s \eq \delta(\sigma_1, 0) \cdots   \delta(\sigma_L, 0) 
    \period \ee
   
   For $M$ large, it follows that
   \be Z \sim ( \lambda_{\rm max} )^{M} \comma \ee
   so we have reduced the calculation of $\kappa$  to the  
   calculation of 
    the maximum eigenvalue  of an $N^L$-dimensional matrix. For the 
    above-mentioned $N =2$, $L = 10$ case  this dimension is 
    $2^{10} = 1024$, which is a huge improvement on $10^{30}$!

{{\color{\ctnclr}
  \section{Ising model}}
\setcounter{equation}{0}


Even so, one still wants to take the limit $L \rightarrow \infty$, so 
$T$ is going to become infinite-dimensional. For the Ising model,
$\sigma_i = \pm 1$ and 
\be \label{Isingwts}
W(\sigma_i,\sigma_j ) \eq \ex^{K \sigma_i \sigma_j} \sep 
\Wb (\sigma_i,\sigma_j ) \eq \ex^{\Kb  \sigma_i \sigma_j}   \period \ee
Onsager considered this model, with the more usual $90^\circ$ 
orientation of the lattice and toroidal (cyclic) boundary conditions.
Then $T$ is the product of two matrices, one of which adds the 
horizontal edges within a row, while the other adds the vertical 
edges between rows. He showed that these two matrices generate a 
finite-dimensional Lie algebra, now known as the Onsager 
algebra.{\color{\ctnclr} \cite[eqns. 60 \& 61]{Onsager1944}}. This 
enabled him
to calculate  {\em all} the eigenvalues of $T$, and hence 
$Z$ and $\kappa$.

The matrix $T$ commutes with the operator $R$ that negates all 
spins:
\be R_{s,s'} \eq \prod_{i=1}^L  \delta(\sigma_i,-\sigma_i') 
\sep R^2 = 1  \ee
so its eigenvectors $v$ either lie in the sub-space where
$Rv = v$ or in  the sub-space $Rv = - v$. In the former case they 
are contained in the eigenvalues of $T_{+}$, in the latter in $T_{-}$,
where
\be T_{\pm} \eq t_1 \otimes t_2 \otimes \cdots \otimes t_L \ee
and the $t_i$ are two-by-two matrices (the $t_i$  are different for
$T_+$ and $T_-$). So in this sense $T$ is a direct product of
$L$ two-by-two matrices.

Bruria Kaufman later gave a simpler derivation of this result, using
anti-commuting spinor (free-fermion) operators, i.e. a Clifford 
algebra.{\color{\ctnclr} \cite{Kaufman1949}}

{{\color{\ctnclr}
\section{Ising model Magnetization} 
\setcounter{equation}{0}

 \subsection{Kaufman and Onsager}}

In 1949, at a conference in Florence, 
Italy,{\color{\ctnclr} \cite{Onsager1949}}
 Onsager referred to 
the magnetization of the Ising model and announced that
``B. Kaufman and I have recently solved'' this problem. He gave 
the result as 
\be  \label{Isingmag}
 {\cal M} \eq (1-k^2)^{1/8} \ee
where
\be \label{defk}
k = 1/(\sinh 2 K   \,  \sinh 2 \Kb )  \ee
and the result is true for $0 <k <1$. For $k >1$ the magnetization 
vanishes, i.e. ${\cal M} = 0 $, so the graph of $\cal M$ is as
in Figure {\color{\ctnclr} \ref{graphM}}.

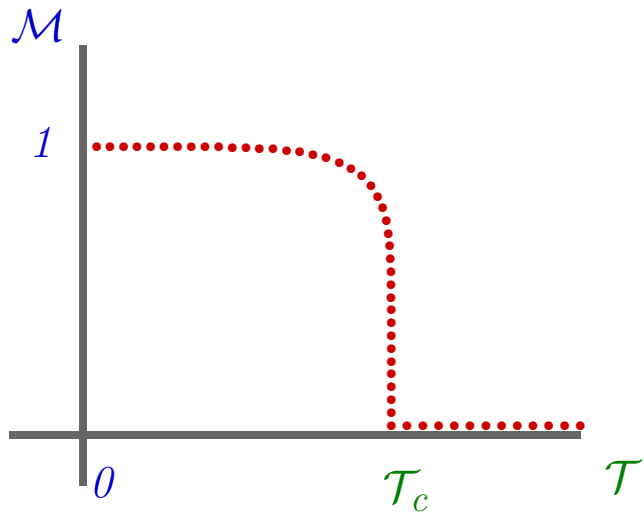
\begin{figure}
\begin{picture}(300,300) (5,-57)
\setlength{\unitlength}{0.9pt}
\thicklines
{\color{black}

\put (110,-35) {\line(0,1) {185}}
\put (111,-35) {\line(0,1) {185}}
\put (112,-35) {\line(0,1) {185}}

\put (80,-15) {\line(1,0) {240}}
\put (80,-14) {\line(1,0) {240}}
\put (80,-13) {\line(1,0) {240}}

{\color{red}

\put(110.0, 105.0)   { \scriptsize $ \bullet $}
\put(115.7, 105.0)   { \scriptsize $ \bullet $}
\put(121.4, 105.0)   { \scriptsize $ \bullet $}
\put(127.1, 105.0)   { \scriptsize $ \bullet $}
\put(132.8, 105.0)   { \scriptsize $ \bullet $}
\put(138.5, 105.0)   { \scriptsize $ \bullet $}
\put(144.2, 105.0)   { \scriptsize $ \bullet $}
\put(149.9, 105.0)   { \scriptsize $ \bullet $}
\put(155.7, 105.0)   { \scriptsize $ \bullet $}
\put(161.4, 105.0)   { \scriptsize $ \bullet $}
\put(167.1, 104.9)   { \scriptsize $ \bullet $}
\put(172.8, 104.7)   { \scriptsize $ \bullet $}
\put(178.5, 104.5)   { \scriptsize $ \bullet $}
\put(184.2, 104.2)   { \scriptsize $ \bullet $}
\put(189.9, 103.7)   { \scriptsize $ \bullet $}
\put(195.5, 102.9)   { \scriptsize $ \bullet $}
\put(201.1, 101.9)   { \scriptsize $ \bullet $}
\put(206.6, 100.5)   { \scriptsize $ \bullet $}
\put(212.0, 98.61)   { \scriptsize $ \bullet $}
\put(217.0, 96.10)   { \scriptsize $ \bullet $}
\put(221.6, 92.87)   { \scriptsize $ \bullet $}
\put(225.4, 88.88)   { \scriptsize $ \bullet $}
\put(228.3, 84.27)   { \scriptsize $ \bullet $}
\put(230.4, 79.27)   { \scriptsize $ \bullet $}
\put(231.8, 74.05)   { \scriptsize $ \bullet $}
\put(232.7, 68.73)   { \scriptsize $ \bullet $}
\put(233.3, 63.37)   { \scriptsize $ \bullet $}
\put(233.6, 58.00)   { \scriptsize $ \bullet $}
\put(233.8, 52.62)   { \scriptsize $ \bullet $}
\put(233.9, 47.23)   { \scriptsize $ \bullet $}
\put(233.9, 41.85)   { \scriptsize $ \bullet $}
\put(234.0, 36.46)   { \scriptsize $ \bullet $}
\put(234.0, 31.08)   { \scriptsize $ \bullet $}
\put(234.0, 25.69)   { \scriptsize $ \bullet $}
\put(234.0, 20.31)   { \scriptsize $ \bullet $}
\put(234.0, 14.92)   { \scriptsize $ \bullet $}
\put(234.0, 9.539)   { \scriptsize $ \bullet $}
\put(234.0, 4.155)   { \scriptsize $ \bullet $}
\put(234.0, -1.230)   { \scriptsize $ \bullet $}
\put(234.0, -6.615)   { \scriptsize $ \bullet $}
\put(234.0, -12.00)   { \scriptsize $ \bullet $}
\put(240.615, -12)   { \scriptsize $ \bullet $}
\put(247.23, -12)   { \scriptsize $ \bullet $}
\put(253.845, -12)   { \scriptsize $ \bullet $}
\put(260.46, -12)   { \scriptsize $ \bullet $}
\put(267.075, -12)   { \scriptsize $ \bullet $}
\put(273.69, -12)   { \scriptsize $ \bullet $}
\put(280.305, -12)   { \scriptsize $ \bullet $}
\put(286.92, -12)   { \scriptsize $ \bullet $}
\put(293.535, -12)   { \scriptsize $ \bullet $}
\put(300.15, -12)   { \scriptsize $ \bullet $}
\put(306.765, -12)   { \scriptsize $ \bullet $}
\put(313.38, -12)   { \scriptsize $ \bullet $}

}

}
{\color{blue}
\put (72,152){\Large  {$\cal   M$}}
  }

{\color{green}
\put (224,-42){\Large  {$\cal  T$}}
\put (231,-46){ \large {\it  c}}
  }

{\color{blue}
\put (71,103){\Large  {\it 1}}
  }

{\color{blue}
\put (92,-40){\Large  {\it 0}}
  }

{\color{green}
\put (304,-38){\Large  {$\cal   T$}}
  }

 \end{picture}
 \caption{\small  ${\cal M}$ as a function of temperature $\cal T$.}
  \label{graphM}
 \end{figure}
 

Onsager and Kaufman did not publish their derivation, but 
we have convincing evidence of their method.
In 1949 they published a 
paper{\color{\ctnclr} \cite{KaufmanOnsager1949}}
 entitled ``Crystal statistics. III"   where they use the 
 free-fermion spinor 
 operators to calculate
the correlation $\langle s_{1,1} s_{i,1+k} \rangle$
for $i= 1,2 $ and any integer $k$ ({\em not} the 
$k$ of {\color{\ctnclr}  (\ref{defk})}) in terms of a 
$k$-dimensional  Toeplitz determinant $\Delta_k$.
The magnetization can be obtained from these
expressions by taking the limit $k \rightarrow \infty$, 
when $\langle s_{1,1} s_{i,1+k} \rangle \rightarrow 
{\cal M}^2 $.


So they needed to calculate the large-size limit of a 
Toeplitz determinant. In a letter dated April 12, 1950, 
Onsager wrote to Kaufman giving a general formula
for this  limit   
$\Delta_{\infty}$.{\color{\ctnclr} \cite{JSP1995}} 
On page {\color{\ctnclr} 3} he says 
``we get the degree of order from C.S.III without much trouble.
It equals $(1 - k^2)^{1/8}$ as before". This letter is also in the 
Onsager archive in Trondheim,  at pages  
{\color{\ctnclr} 21 -- 24}   of 

\noindent {\footnotesize 
{\color{\ctnclr}  {http://www.ntnu.no/ub/spesialsamlingene/
$\! \! $tekark/tek5/research/009{\_}0097.html}}}

\noindent In the same archive,  on pages 
{\color{\ctnclr} 32, 33} of 

\noindent  {\footnotesize  
{\color{\ctnclr}  {http://www.ntnu.no/ub/spesialsamlingene/
$\! \! $tekark/tek5/research/009{\_}0096.html}}}

\noindent there is a letter from Kaufman to Onsager
thanking him for his letter and saying she has worked out a way of 
using long-range order along a row, rather than a diagonal, 
and applying his procedure to obtain $\Delta_{\infty}$. She 
goes on to say that the mathematician Kakutani had written to her
saying that he had spoken to Onsager about this, and 
was very interested.

Onsager recounts something of what happened in 1971. In 
{\color{\ctnclr} \cite{Onsager1971a}}   he says that  Kaufman
derived the correlations  in the form of recurrent (Toeplitz)
determinants, and they were particularly simple along a diagonal
(i.e. a row or column of Figure {\color{\ctnclr}  \ref{sqlatt}}). He 
then discusses
the problem of calculating the infinite-size limit of these 
determinants and indicates how he first did this for the 
particular problem by determining the
eigenvalues and taking their product. This was the basis for 
his announcement of the result in Florence.

He then looked for a more general formula for Toeplitz 
determinants with arbitrary generating functions. He found 
one for a large class of rational generating functions. 


From this, he says ``the general result stared me in the face". 
This is the formula {\color{\ctnclr} (7)}
 of  {\color{\ctnclr} \cite{JSP1995}}, which 
is exact for for his rational generating functions  when
the dimension $k$ of the determinant is finite but sufficiently 
large. Since  the elements of $\Delta_{k}$ are the central
$2 k -1$ Laurent coefficients of the generating function,
in the limit $k \rightarrow \infty$  it  should be  possible to
extend {\color{\ctnclr} (7)} to more arbitrary functions, providing 
appropriate conditions for convergence are fulfilled.
His last sentence
reads ``Only, before I knew what sort of conditions to impose on 
the generating function, we talked to Kakutani and Kakutani 
talked to Szeg{\H o}  and the mathematicians got there first."

Further explanation of that comment is given 
in  {\color{\ctnclr} \cite{Onsager1971b}}, 
 where he says that he had found 
``a general formula for the evaluation of Toeplitz 
matrices.\footnote{Refs. {\color{\ctnclr} \cite{Onsager1971a}},
{\color{\ctnclr} \cite{Onsager1971b}} are reprinted in 
Onsager's collected works, pages {\color{\ctnclr} 232 -- 241}
  and {\color{\ctnclr}37 -- 45}, 
respectively.{\color{\ctnclr} \cite{Onsager1996}} }
 The only thing I did not know was how to fill out the holes
 in the mathematics and show the epsilons and the deltas 
 and all of that''.  Onsager adds  that six years later
 the mathematician Hirschman told him that he could
 readily have  completed his proof by using
 a theorem of  Wiener's.

Szeg{\H o} did publish his resulting
general theorem{\color{\ctnclr} \cite{Szego1952},
\cite{Kac1954}}  
on the large-size limit of a Toeplitz 
determinant, but not until 1952.

So Onsager had a derivation of $\cal M$ in 1949, but looked for
a more general way to calculate $\Delta_{\infty}$. For his rational 
generating functions he actually proved what is now 
Szeg{\H o}'s theorem. He could see the extension to more 
general functions, and that it gave the same result
{\color{\ctnclr} (\ref{Isingmag})} as his previous method. 
However, he lacked a 
rigorous proof of that extension and did not pursue
the matter when the mathematicians became interested.

The author also has a copy of a typescript, given to him by John 
Stephenson,  entitled   ``Long-Range Order"  and with  three 
names hand-written at the top: ``Onsager'' nearest the title,
``B. KAUFMAN'' above that, and ``R. B. Potts" above that. It 
shows how  the magnetization can be obtained from eqn. 
{\color{\ctnclr}  43} of {\color{\ctnclr}  \cite{KaufmanOnsager1949}}, 
using Onsager's  working  of 
{\color{\ctnclr}  \cite{JSP1995}} and deriving the formula of 
the above-mentioned 
letter from  Kaufman to Onsager. This 
must be a draft of a paper by Onsager and Kaufman
on the subject.  The author intends to make this typescript 
available at {\color{\ctnclr}  $<$http://arxiv.org/$>$}.

\pagebreak

{{\color{\ctnclr}
\subsection{Yang and others}}
The first derivation of $\cal M$ published  was in 1952 by  C.N. 
Yang.{\color{\ctnclr} \cite{Yang1952}}  He first wrote $\cal M$
in terms of the two maximal eigenvectors of the transfer matrix. 
He then used Kaufman's spinor operators
to write $\cal M$ as the product of eigenvalues
(i.e. a determinant) of an $L$ by $L$ matrix and evaluated 
 the eigenvalues in the limit $L \rightarrow \infty $.

Later, combinatorial ways were found of writing the partition function 
of the Ising model on a finite lattice directly as  a determinant or a 
pfaffian (the square root of an anti-symmetric
determinant).{\color{\ctnclr} \cite{KacWard, HurstGreen}}. Then
 it was
realised that the problem could be solved by first expressing it 
as one of filling a planar lattice with dimers.{\color{\ctnclr} 
\cite{Kasteleyn1961, Fisher,TemperleyFisher, Kasteleyn1963}}
In 1963 Montroll, Potts and Ward{\color{\ctnclr} \cite{MPW1963}} 
showed that 
the magnetization could be written as the ratio of two determinants, 
thereby obtaining a combinatorial proof of the 
result {\color{\ctnclr}  (\ref{Isingmag})}. 
The numerator determinant is indeed  Toeplitz and they  evaluated 
its large-size limit  by using  Szeg{\H o}'s  theorem.






{{\color{\ctnclr}
\vspace{3mm}
 \section{Six-vertex model}}
  \setcounter{equation}{0}

 For any $N$-state model (with nearest neighbour interactions) on a 
 lattice of   $L$ columns, the transfer matrix $T$  is of dimension 
 $N^L$.  So is the hamiltonian $H$  of a quantum mechanical system
 of   $N$-state spins on a line of $L$ sites. Calculating the largest 
 eigenvalue of $T$ corresponds to calculating the lowest energy 
 state of $H$.

In 1966 C.N. Yang and C.P. Yang{\color{\ctnclr} \cite{Yang1966}} 
extended the Bethe ansatz by using it to calculate the 
ground-state energy of the hamiltonian of the anisotropic Heisenberg
chain, also known as the XXZ chain. The following year Elliott Lieb 
used this method to calculate $\kappa$
for three particular models: ice, F and 
KDP.{\color{\ctnclr} \cite{Lieb1,Lieb2,Lieb3}}
These are all special cases of a more general zero-field ``ice-type" 
or  ``six-vertex" model.  
The solution of this model was given in 
the same year by  Sutherland.{\color{\ctnclr} \cite{Sutherland1967}}
Finally, the solution of the general six-vertex model was given 
by C. P. Yang{\color{\ctnclr}\cite{YangCP1967}} and  
Sutherland {\it et al} {\color{\ctnclr}\cite{SYY1967}}.

In these models one places arrows on the edges of the square lattice
so that at each vertex there are two arrows in and two arrows out
(this is known as the ``ice rule''). There are six ways of doing this, as 
shown in the upper diagram in Figure
 {\color{\ctnclr}  \ref{sixvertex}}. In general
we assign to these six vertex configurations  the weights
$\omega_1, \ldots  ,  \omega_6$, as in the figure.


 \begin{figure}[hbt]
\begin{picture}(320,180) (-34,-20)

\setlength{\unitlength}{1.0pt}

\put(32,53){$\omega_1$}
\put(88,53){$\omega_2$}
\put(144,53){$\omega_3$}
\put(200,53){$\omega_4$}
\put(256,53){$\omega_5$}
\put(312,53){$\omega_6$}

{\color{blue}

\put (10,100) {\line(1,0) {44}}
\put (66,100) {\line(1,0) {44}}
\put (122,100) {\line(1,0) {44}}
\put (178,100) {\line(1,0) {44}}
\put (234,100) {\line(1,0) {44}}
\put (290,100) {\line(1,0) {44}}

\put (32,78) {\line(0,1) {44}}
\put (88,78) {\line(0,1) {44}}
\put (144,78) {\line(0,1) {44}}
\put (200,78) {\line(0,1) {44}}
\put (256,78) {\line(0,1) {44}}
\put (312,78) {\line(0,1) {44}}

\put (10,100.5) {\line(1,0) {44}}
\put (66,100.5) {\line(1,0) {44}}
\put (122,100.5) {\line(1,0) {44}}
\put (178,100.5) {\line(1,0) {44}}
\put (234,100.5) {\line(1,0) {44}}
\put (290,100.5) {\line(1,0) {44}}

\put (32.5,78) {\line(0,1) {44}}
\put (88.5,78) {\line(0,1) {44}}
\put (144.5,78) {\line(0,1) {44}}
\put (200.5,78) {\line(0,1) {44}}
\put (256.5,78) {\line(0,1) {44}}
\put (312.5,78) {\line(0,1) {44}}

\put (28.5,110) {$\wedge$}
\put (28.5,86) {$\wedge$}

\put (84.5,111) {$\vee$}
\put (84.5,85) {$\vee$}

\put (15,97.5) {$>$}
\put (41,97.5) {$>$}

\put (71,97.5) {$<$}
\put (97,97.5) {$<$}

\put (127,97.5) {$>$}
\put (153,97.5) {$>$}
\put (140.5,111) {$\vee$}
\put (140.5,85) {$\vee$}

\put (183,97.5) {$<$}
\put (209,97) {$<$}
\put (196.5,110) {$\wedge$}
\put (196.5,86) {$\wedge$}

\put (239,97.5) {$>$}
\put (265,97.5) {$<$}
\put (252.5,110) {$\wedge$}
\put (252.5,85) {$\vee$}

\put (295,97.5) {$<$}
\put (321,97.5) {$>$}
\put (308.5,111) {$\vee$}
\put (308.5,86) {$\wedge$}


\multiput(11,12)(3,0){14}{.}
\multiput(32,-9)(0,3){14}{.}



\multiput(123,12)(3,0){14}{.}


\multiput(198,-9)(0,3){14}{.}


\multiput(254,13)(0,3){7}{.}

\multiput(235,12)(3,0){7}{.}


\multiput(311,-9)(0,3){7}{.}

\multiput(312,12)(3,0){7}{.}

}
{\color{red}



\put (62,13) {\line(1,0) {21}}
\put (62,13.5) {\line(1,0) {21}}
\put (62,14) {\line(1,0) {21}}

\put (85,11) {\line(1,0) {22}}
\put (85,11.5) {\line(1,0) {22}}
\put (85,12) {\line(1,0) {22}}

\put (85,-10) {\line(0,1) {21}}
\put (85.5,-10) {\line(0,1) {21}}
\put (86,-10) {\line(0,1) {21}}

\put (83,13) {\line(0,1) {21}}
\put (83.5,13) {\line(0,1) {21}}
\put (84,13) {\line(0,1) {21}}


\put (140,-10) {\line(0,1) {44}}
\put (140.5,-10) {\line(0,1) {44}}
\put (141,-10) {\line(0,1) {44}}


\put (174,12) {\line(1,0) {44}}
\put (174,12.5) {\line(1,0) {44}}
\put (174,13) {\line(1,0) {44}}


\put (252,12) {\line(1,0) {22}}
\put (252,12.5) {\line(1,0) {22}}
\put (252,13) {\line(1,0) {22}}

\put (252,-10) {\line(0,1) {22}}
\put (252.5,-10) {\line(0,1) {22}}
\put (253,-10) {\line(0,1) {22}}


\put (286,12) {\line(1,0) {22}}
\put (286,12.5) {\line(1,0) {22}}
\put (286,13) {\line(1,0) {22}}

\put (308,12) {\line(0,1) {22}}
\put (308.5,12) {\line(0,1) {22}}
\put (309,12) {\line(0,1) {22}}

}

\end{picture}

 \caption{\small The six arrangements of arrows at a vertex
 in the six-vertex model .} 
  \label{sixvertex}
  \vspace{1cm}

 \end{figure}
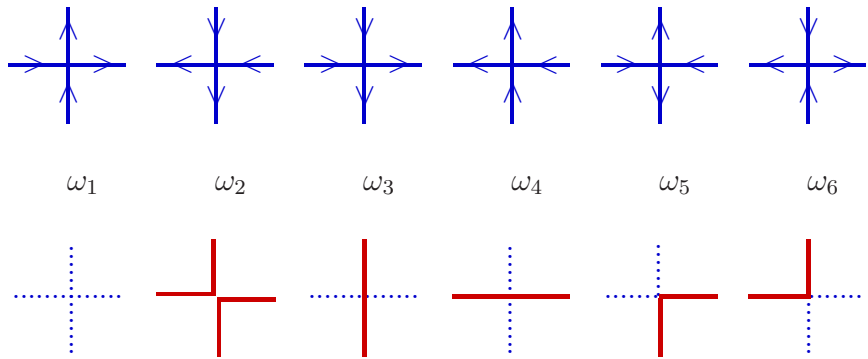

One can place a line on any horizontal edge
bearing a left-pointing arrow and a line on  any vertical edge
bearing a down-pointing arrow; other edges one regards
as empty.
The six possible vertex configurations  are then represented 
as in the lower diagram of Figure 
{\color{\ctnclr}  \ref{sixvertex}}, using dots 
 to represent empty edges.
Separating the lines in the second vertex as indicated, it 
becomes 
apparent that these lines are continuous, and can be viewed as 
moving generally up and to the right through the lattice.


If we impose cyclic boundary conditions, then it follows that 
the number of lines in one row of vertical edges must be 
the same as the number in the row above. (Lines can move out
at the right-hand boundary, but must re-appear at the left.)
Hence if there are $n$ lines in one row, then there are $n$ 
lines in all rows. This is the ``line-conservation''  
or ``conservation of down arrows'' property. 
The row-to-row transfer matrix $T$ therefore breaks up into 
$L+1$ diagonal blocks, one for each possible value
 ($0, 1, \ldots, L$) of $n$.
Two typical arrangements
of lines in two  adjacent rows are shown in Figure
{\color{\ctnclr}  \ref{rows}}.


 \begin{figure}[hbt]
\begin{picture}(320,180) (-24,-20)

\setlength{\unitlength}{1.0pt}

{\color{blue}

\multiput(10,19)(3,0){50}{.}
\multiput(210,19)(3,0){50}{.}

\put (30,-14) {$x_1$}
\put (90,-14) {$x_2$}
\put (260,-14) {$x_1$}
\put (327,-14) {$x_2$}

\put (62,50) {$y_1$}
\put (112,50) {$y_2$}
\put (239,50) {$y_1$}
\put (294,50) {$y_2$}

}

{\color{red}


\put (29.3,20) {\line(1,0) {30}}
\put (29.3,20.5) {\line(1,0) {30}}
\put (29.3,21) {\line(1,0) {30}}

\put (29.5,-2) {\line(0,1) {22}}
\put (30,-2) {\line(0,1) {22}}
\put (30.5,-2) {\line(0,1) {22}}

\put (59.5,20) {\line(0,1) {22}}
\put (60,20) {\line(0,1) {22}}
\put (60.5,20) {\line(0,1) {22}}


\put (89.3,20) {\line(1,0) {20}}
\put (89.3,20.5) {\line(1,0) {20}}
\put (89.3,21) {\line(1,0) {20}}

\put (89.5,-2) {\line(0,1) {22}}
\put (90,-2) {\line(0,1) {22}}
\put (90.5,-2) {\line(0,1) {22}}

\put (109.5,20) {\line(0,1) {22}}
\put (110,20) {\line(0,1) {22}}
\put (110.5,20) {\line(0,1) {22}}


\put (207,20) {\line(1,0) {30}}
\put (207,20.5) {\line(1,0) {30}}
\put (207,21) {\line(1,0) {30}}

\put (236.5,20) {\line(0,1) {22}}
\put (237,20) {\line(0,1) {22}}
\put (237.5,20) {\line(0,1) {22}}


\put (259.3,20) {\line(1,0) {32}}
\put (259.3,20.5) {\line(1,0) {32}}
\put (259.3,21) {\line(1,0) {32}}

\put (259.5,-2) {\line(0,1) {22}}
\put (260,-2) {\line(0,1) {22}}
\put (260.5,-2) {\line(0,1) {22}}

\put (291.5,20) {\line(0,1) {22}}
\put (292,20) {\line(0,1) {22}}
\put (292.5,20) {\line(0,1) {22}}


\put (326.3,20) {\line(1,0) {30}}
\put (326.3,20.5) {\line(1,0) {30}}
\put (326.3,21) {\line(1,0) {30}}

\put (326.5,-2) {\line(0,1) {22}}
\put (327,-2) {\line(0,1) {22}}
\put (327.5,-2) {\line(0,1) {22}}

}

\end{picture}

 \caption{\small The two typical arrangements of 
lines in adjacent rows of the six-vertex model. The 
$y_1, \ldots y_n$ must interlace the $x_1, \ldots ,x_n$. } 
  \label{rows}
  \vspace{1cm}

 \end{figure}
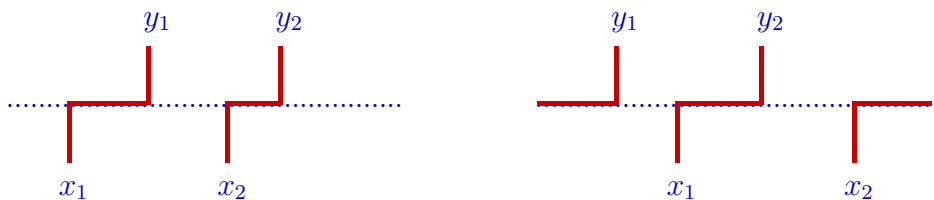

{{\color{\ctnclr}
 \subsection{The Bethe ansatz} }
 Here I shall briefly outline how the Bethe ansatz is applied
 to the usual row-to-row transfer matrix. In some ways it is simpler
 if one instead uses the transfer matrix in the diagonal 
 SW to NE direction, since then lines only move at most 
 one position at a time. Alternatively, helical boundary 
 {\color{\ctnclr}\cite{Baxter1969}} conditions mean that one only 
 has to add a single vertex at a time. 
 
 However, it is worth the effort to work with the 
 row-to-row matrix, since the final equations simplify and 
 can be compared with hamiltonian calculations such 
 those mentioned above by Yang. Further, we shall find that
 they manifest the important commutation property of 
 the transfer matrices.
 
 The method is discussed in  {\color{\ctnclr}\cite{LiebWu,Gaudin}}
 and, for the zero-field case  (when $\omega_1 = \om_2$ and $ 
 \om_3 = \om_4$), in sections {\color{\ctnclr} 8.2 - 8.4} of
{\color{\ctnclr} \cite{book}}. Here I outline the extension
to arbitrary $\om_1, \ldots, \om_6$.

Let $\Lambda$ be an eigenvalue of the transfer matrix $T$
and $f$ the corresponding eigenvector. Each element
of $f$ is associated with a configuration of arrows or lines
on a row of vertical edges of the lattice. Write
$f(x_1,\ldots, x_n)$ for the element   
corresponding to lines (down arrows) 
 at positions
$x_1, \ldots ,x_n$ be $f(x_1,\ldots, x_n)$. Write the 
set $x_1, \ldots x_n$ as $X$, and $y_1, \ldots y_n$ as $Y$.
We must have
\be 1  \leq x_1  < x_2 < \cdots   x_L \leq L \ee
and similarly for $Y$.


 Then the eigenvalue
equation is
 \be \label{eigval}
 \Lambda f(X) \eq \sum C(X,Y) f(Y) +  \sum D(X,Y) f(Y) \comma \ee
 where the first sum is over all $y_1, \ldots , y_n$ such that
 \bd x_1 \leq y_1 \leq x_2 \leq  y_2  \leq  \ldots   \leq x_n \leq 
 y_n  \leq L \comma \ed
  and the second sum is over 
 \bd   1 \leq y_1 \leq  x_1 \leq y_2  \leq  x_2 \leq \ldots   \leq y_n 
 \leq x_n \period \ed
 
 Here $X$ is the configuration of arrows in one row, and $Y$ is
 the configuration in the row above, as in Figure 
  {\color{\ctnclr}  \ref{rows}}.
 The coefficients $C(X,Y), D(X,Y)$ are the products of the 
 Boltzmann weights of the vertices  in between. Note that
 special cases occur when an $x_i$ is equal to a $y_j$, as then
 we can get a vertex of type 2 or type 3, with weight 
 $\omega_2$ or $\omega_3$. We must also ensure that we do 
 not include cases where two of the $y_i$ are equal,
 e.g. $y_1 = x_2 = y_2$ in the first sum.

{{\color{\ctnclr}
\subsection{The case $n=1$}} 
When $n= 1$ the 
 equation {\color{\ctnclr}  (\ref{eigval})} is (writing 
 $x_1, y_1$ simply as $x,y$)
 \ba  \label{nis1}
 \Lambda f(x)  & = &
  \om_1^{L-1}\om_3 f(x) +
 \om_5  \, \om_6 \sum_{y=x+1}^L   \om_1^{L+x-y-1}   \om_4^{y-x-1} f(y) 
 \nonumber \\
 &&  + \om_5 \, \om_6 \sum_{y=1}^{x-1} 
   \om_1^{x-y-1}   \om_4^{L+y-x-1} f(y)  + \om_2 \, \om_4^{L-1} f(x) 
   \period \ea
 We try the solution
 \bd f(x) = z^x \ed
 and find the RHS of  {\color{\ctnclr} (\ref{nis1})}
  is
 \be  \displaystyle (\om_1^L {\cal L} +\om_4^L {\cal M})z^x
 + \rho (z) \,  \om_1^{x} \, \om_4^{L-x} (1-z^L)  \comma \ee
 where \ba  \label{defLM}
 \rho (z)  = &   \!  \!   \!  \!   \!  \!   \!  \!   \!  \!  
  \!  \!   \!  \!   \!  \!   \!  \!   \!  \!    {\displaystyle \frac{\om_5 \om_6  z}
{ \om_1 (\om_1 - \om_4 z ) }  } \nonumber \\
 {\cal L}  = {\cal L}(z) = &   \!  \!  {\displaystyle  \frac{\om_1 \, \om_3 +
  (\om_5 \, \om_6 - \om_3 \, \om_4 )z}
  {\om_1 (\om_1-\om_4 z) }   } \\
   {\cal M}   = {\cal M}(z) = &  \! \!  
    { \displaystyle  \frac{\om_1 \, \om_2- \om_5 \, \om_6  
   -\om_2 \, \om_4 z}
  {\om_4 (\om_1-\om_4 z) } } \period  \nonumber \ea
  Note that $\cal M$ is this section is {\em not} the
 same as the magnetization $\cal M$ discussed elsewhere
 in this paper.
It follows that  {\color{\ctnclr} (\ref{nis1})} is satisfied if
  \bd \Lambda = \om_1^L {\cal L}(z)  +\om_4^L {\cal M}(z) 
  \; \; \; {\rm and } \; \; \;   z^L = 1 \period \ed
  Thus there are $L$ possible values for
  $z$ (all lying on the unit circle),
  giving the  $L$ eigenvalues $\Lambda$ in the $n=1$ 
  block of the transfer matrix.
  

{{\color{\ctnclr}
 \subsection{\bf The case $n=2$}}
   When $n= 2$  we first try
    \bd f(x_1,x_2 ) = z_1^{x_1} \, z_2^{x_2} \period  \ed
Then  various terms come from  the sums represented by the
two diagrams in Figure {\color{\ctnclr}  \ref{rows}}. Considered as
 functions of $x_1, x_2$, the only ones proportional to 
 $ z_1^{x_1} \, z_2^{x_2} $ are
 \bd \left[ \om_1^L {\cal L} (z_1)  {\cal L} (z_2)  +\om_4^L 
  {\cal M} (z_1)  {\cal M}  (z_2)  \right]  z_1^{x_1} \, z_2^{x_2} 
  \comma \ed
  so these terms in the eigenvalue equation are satisfied if
    \bd \Lambda =  \om_1^L {\cal L} (z_1)  {\cal L} (z_2)  +\om_4^L 
  {\cal M} (z_1)  {\cal M}  (z_2)  \period \ed
  
  We shall see that the contributions from the boundary terms
  lead to the ``conservation of momentum'' equation
  \be \label{momntm}
   z_1^L z_2^L = 1 \period \ee
  Both these last two ``energy'' and ``momentum'' equations
  are unchanged by permuting $z_1$ and $z_2$,  so we could more
  generally try an ansatz of the form
   \bd f(x_1,x_2 ) = A_{12} z_1^{x_1} \, z_2^{x_2} + 
   A_{21} z_2^{x_1} \, z_1^{x_2} \comma  \ed
   where $A_{12}$  and  $A_{21}$ are coefficients that are
    at our disposal.
   Then the boundary terms contain factors
   $\rho (z_1)$ or $\rho (z_2)$ and vanish if
   \bd A_{12} = z_1^L A _{21} \sep A_{21} = z_2^L A _{12} 
   \comma \ed
   from which {\color{\ctnclr} (\ref{momntm})} follows.
   
   The other terms that arise are ``unwanted internal terms''
    terms from the first diagram
   of Figure {\color{\ctnclr}   {\ref{rows}}}
    containing a factor   $(z_1 z_2)^{x_2}$
   and from the second diagram 
    containing a factor   $(z_1 z_2)^{x_1}$. Both of these
    vanish
    if
    \be \label{sA}  s(z_1, z_2) A_{12} + s(z_2,z_1) A_{21} = 0 
    \comma \ee
    where  $s(z_1,z_2)  =   \om_1 \om_4 {\cal M}(z_1) {\cal L}(z_2) -
     \om_2 \om_3 $. Removing factors that cancel out of 
    {\color{\ctnclr}   (\ref{sA})}, we can take
     \be \label{defs}
      s(z_1,z_2)  = \frac{\om_1 \om_3 - (\om_1 \om_2 +
      \om_3 \om_4 - \om_5 \om_6 ) z_2 + 
     \om_2 \om_4 z_1 z_2}{\om_1 \om_3 } \period \ee

{{\color{\ctnclr}
\subsection{Arbitrary $n$} }
      It turns out that these results generalise easily to arbitrary $n$, 
 becoming
 \be \label{eigvec}
 f(X) \eq \sum_P A_{P} z_{p_1} ^{x_1}  \cdots  z_{p_n} ^{x_n} 
 \comma \ee
 where the sum is over all $n!$ permutations 
 $P = \{ p_1, p_2, \ldots , p_n \} $ of the integers $\{ 1, \ldots , n \} $
 and  $A_P = A_{p_1, \ldots ,p_n}$.
 This is the Bethe ansatz. The unwanted internal terms give the
  equations
  \be \label{sPA}
  s(z_{p_j}, z_{p_{j+1}} ) A_P +  
  s(z_{p_{j+1}}, z_{p_{j}} ) A_{P^{(j)}} \sep   1 \leq j < n \ee
  where $P^{(j)} $ differs from $P$ in the interchange 
  of $p_j$ with $p_{j+1}$.  There are many more equations than 
  unknowns in {\color{\ctnclr} (\ref{sPA})}, 
  but they have the solution
  \be A_P \eq \epsilon_P \prod_{1 \leq i < j \leq n } s(z_{p_j},z_{p_i} )
  \comma    \ee
  where   $ \epsilon_P = +1$ for even permutations, $-1$ 
  for odd permutations.

  
  The boundary terms give
  \be A_P \eq z_{p_1}^L A_{ p_2,  \ldots , p_{n}, p_1} \ee
  Eliminating all $A_P$ between these two equations, we get
  the $n$  ``Bethe equations''
  \be \label{Betheeqns}
  z_j^L \eq (-1)^{n-1} \prod_{m=1, m  \neq j} ^n 
  \frac{s(z_m,z_j)}{s(z_j,z_m)} 
  \period \ee These imply $z_1^L \cdots z_n^L = 1$
  and in general define $z_1, \ldots ,z_L$. There are many 
  solutions, corresponding to the different eigenvalues.
  Finally, the wanted terms in the eigenvector equation
  give the eigenvalue as
  \be \label{Lambda}
   \Lambda \eq \om_1^L {\cal L} (z_1) \cdots {\cal L} (z_n)
  + \om_4^L  {\cal M} (z_1) \cdots {\cal M} (z_n) \period \ee
  
  For the ice model, where $\om_1 = \cdots = \om_6$, the full 
  working  is given in Ref. {\color{\ctnclr}\cite{Lieb1}}

{{\color{\ctnclr}
\subsection{Free-fermion  case} }

Note that when
\bd  \om_1 \om_2 +
      \om_3 \om_4 - \om_5 \om_6   =  0 \ed
the function $s(z_1,z_2)$ in {\color{\ctnclr}  (\ref{defs})}
 is symmetric, so 
{\color{\ctnclr}  (\ref{Betheeqns})} simplifies to
\bd z_j^L = (-1)^{n-1}    \comma  \ed
and the Bethe equations  can be solved explicitly for 
all the eigenvalues of $T$.

This is known as the ``free-fermion''  case. LIke the Ising model, 
it can also be solved combinatorially by 
pfaffians.{\color{\ctnclr}\cite{FanWu}}

{{\color{\ctnclr} \subsection{Zero-field case}}
  The zero-field case is when
  \be \om_1 = \om_2 = a \sep \om_3 = \om_4 = b \sep 
  \om_5 = \om_6 = c \comma \ee
  $a,b,c$ being positive parameters.  The vertices 5 and 6 
  always occur in pairs, being  sinks and sources of horizontal 
  arrows, so there is no restriction in taking
  $\om_5 = \om_6$.  The first two conditions are definitely 
  constraints. Then from {\color{\ctnclr}  (\ref{defs})}, 
  \be \label{smj}
  s(z_m,z_j) \eq 1- 2  \Delta z_j + z_m z_j \comma \ee
  where
  \be \label{defD}
  \Delta = \frac{a^2+b^2-c^2}{2 a b } \period \ee

It turns out to be useful to express these equations in terms of 
new variables $\mu, v, \beta_j$ defined by
\bd a = \sin v \, , \; \; b = \sin ( \mu-v) \, ,  \; \; c =  \sin \mu\, , \; \; \ed
\bd  z_j \eq \frac{\sin \beta_j}{\sin ( \mu - \beta_j ) }  \period \ed
This fixes the normalization of $a,b,c$ and ensures that 
\bd \Delta = -\cos \mu   \sep 
{\cal L}(z_j)  = -  \frac{\sin (v-  \beta_j - \mu)}{\sin ( v- \beta_j) }
 \sep 
{\cal M}(z_j)  = -  \frac{\sin ( v- \beta_j + \mu)}{\sin (v - \beta_j) }
 \period   \ed
{From} {\color{\ctnclr}  (\ref{smj})},
\bd s(z_m, z_j ) \eq  G_{mj} \sin(\beta_j - \beta_m + \mu )  \ed
where $G_{mj}  = G_{jm}  = \sin \mu /[
\sin (\mu-\beta_j) \sin(\mu - \beta_m) ]$.
The factors $G_{mj}$ cancel out of  the RHS of 
{\color{\ctnclr}  (\ref{Betheeqns})}, 
so  it depends on 
$\beta_1, \ldots \beta_n$ only via their differences.


If we define
\be \phi (v) \eq \sin^{L}  v  \sep 
q(v) = \prod_{j=1}^n \sin(v- \beta_j) 
\comma \ee
and note that $\Lambda$ is a function of $v$, so write it as $t(v)$, 
then {\color{\ctnclr} (\ref{Betheeqns})} and 
{\color{\ctnclr} (\ref{Lambda})} become
 (for $j = 1, \ldots , n$)
\be \phi(\beta_j) \, q(\beta_j-\mu') + \phi(\mu-\beta_j) \,
q(\beta_j + \mu') \eq 0 \ee

\be \label{tq}
t(v) q(v) \eq  \phi(v)\,  q(v-\mu') + \phi(\mu-v) \, q(v+ \mu')  \ee
where $\mu' = \mu-\pi$. 
When $-1 < \Delta < 1$, we can choose $\mu, v$ to be positive, with  
 $0 <  \mu  <  \pi $, $ 0 <  v < \mu$.


  For $\Delta <1$ and $L$ even, the maximum eigenvalue lies 
  in the block   with 
  \bd n = L/2 \ed
  and the $z_1, \ldots , z_n$ lie on the unit circle, distributed about 
  the point $z=1$. In the thermodynamic  limit of $L$ large they 
  form a continuous distribution and  
  {\color{\ctnclr}  (\ref{Betheeqns})} becomes 
  a linear integral equation for the distribution function. If we 
  transform   from the $z_j$ to the $\beta_j$  (the $\beta_j $ 
  lie on  the vertical line $Re (\beta_j) = \mu/2 $ in the 
  complex  plane), then this equation
  has a difference kernel and the integral is over 
  a full period of the functions, so the equation can be solved
  by Fourier transforms, giving an exact expression for the
  free energy $\log \kappa$. For the $F$ model, Lieb 
  found{\color{\ctnclr}\cite{Lieb2}}  that
   there was a transition of infinite order   at   $\Delta = -1$, i.e. 
   all derivatives of the free energy exist but there is an
   essential  singularity
   of type $\ex^{-C/|{\cal T} -{\cal T}_c |^{1/2}}$, 
   $\cal T $ being the temperature,
   $T_c$ the critical temperature   and $C$ a constant.
   For the KDP model he found{\color{\ctnclr}\cite{Lieb3}} there 
   was a first-order transition at    $\Delta = 1$ from a disordered 
phase ($\Delta<1$),  to 
 a frozen ordered state   ($\Delta>1$)  where the vertical 
   (horizontal) arrows  all point  in the same direction. This 
   was very different    behaviour
   from that of the Ising model, which has a logarithmic
   singularity in the specific heat.
   
{{\color{\ctnclr} 
 \subsection{The model in a field}}
   The above working remains valid when electric fields are
 applied, making $\om_1 \neq \om_2$ and/or  
 $\om_3 \neq \om_4$. The system is then either frozen
 or the maximum eigenvalue occurs when $z_1, ... z_n$ are
 distributed along a curve in the complex plane.
The resulting integral equation cannot in general be solved 
analytically by  Fourier transforms. However, it can be solved 
numerically and at (and sometimes about) various special cases 
(including the free-fermion  case).{\color{\ctnclr} \cite{LiebWu}, 
\cite[\S8.12]{book}, \cite{Jayaprakash1983,
Nolden1992,  Bukman1995, Sato1995}}


{{\color{\ctnclr} 
\vspace{2mm}
\subsection{Matrix functional equations}  }

I had the good fortune to work with Elliott Lieb at the Massachusetts
Institute of Technology from 1968 to 1970, and looked at a number of 
Bethe ansatz problems. In particular I considered the most general
inhomogeneous six-vertex model that could be solved by the Bethe
ansatz, using equations of the same form, but involving different 
parameters at different sites of the 
lattice.{\color{\ctnclr} \cite{BaxterMIT}} In particular, instead of the 
simple exponentials
$z_j^x$ above, one uses a more general ``single-particle''
function  $z_j(x)$. 


In 1970 my wife Elizabeth and I left MIT to return to Australia via 
England,  where we would spent time with my parents in Essex. 
We intended to make the  journey from England to Australia by ship. 
Already it was unusual to do this rather than travel  by air (a change
 that had occurred only in the previous decade). We had the choice 
 of two ships, giving us either two or five  months'  break in England.

We chose five months, and it was a good decision. Towards the 
end of  that time I looked again at my MIT paper and realised 
that there were 
parameters in the final definition of the model that entered the 
eigenvalues of $T$, but {\em not} the eigenvectors. Indeed, this was 
apparent from Lieb and Sutherland's work on the zero-field 
six-vertex model: from {\color{\ctnclr}  (\ref{eigvec}) - (\ref{defD})},
the  eigenvectors $f$ depend only on the single parameter 
$\Delta = - \cos \mu $.
They are the same eigenvectors as those found by Yang and
 Yang{\color{\ctnclr} \cite{Yang1966}}  for  the  hamiltonian of the 
anisotropic Heisenberg chain.

The Boltzmann weights of the model depend  also on $v$,
If we write the transfer matrix $T$ as $T(v)$ and keep $\Delta$ 
and $\mu$ fixed,   then  under quite 
general conditions it follows that two transfer 
matrices  $T(u)$,  $T(v)$, with the same value of $\Delta$,  
{\em commute}:
\be T(u) \, T(v) \eq T(v) \, T(u) \period \ee
There is a common eigenvector matrix $\cal P$, independent of $v$,
 such that $ {\cal P}^{-1} T(v) {\cal P}$ is diagonal, its entries
 being the eigenvalues $t(v)$.

Form a matrix $Q(v)$
such  that $ {\cal P}^{-1} Q(v) {\cal P}$ is also diagonal, with entries
the corresponding $q(v)$ of  {\color{\ctnclr}  (\ref{tq})}. Then
 $T(v)$ and $Q(v)$  must also satisfy  
{\color{\ctnclr}  (\ref{tq})}, i.e. 
\be \label{TQ}
T(v) Q(v) \eq  \phi(v)\,  Q(v-\mu') + \phi(\mu-v) \, Q(v+ \mu') \comma  \ee
the $\phi$ factors being scalars;  $Q(v)$ must commute with
$T(u)$, for all $u,v$.

  
  This matrix relation, together with some elementary 
  observations, is equivalent to the Bethe ansatz. Together with the
  commutation properties, it implies {\color{\ctnclr}  (\ref{tq})}.
   Since the elements of 
  $T$ are   sums of products of $L$ Boltzmann weights, each equal
   to $a,b$ or $c$,  and vertices 5 and 6 always occur in pairs,
    each element must be of the form  $\ex^{-iLv}$ times a 
   polynomial of degree $L$ in $\ex^{2iv}$. If 
   the elements of $Q$ have a similar structure, but with $L$ 
   replaced by $n$, 
   then    the same must be true of the eigenvalues $t(v), q(v)$; 
   and $q(v)$ must have $n$ zeros  $\beta_1, \ldots, \beta_n$.
   Setting $v = \beta_j$ in {\color{\ctnclr}  (\ref{tq})} 
   (and noting that $t(v)$ is entire, 
   so finite), we obtain the Bethe equations 
   {\color{\ctnclr} (\ref{Betheeqns})}.
   
 
{{\color{\ctnclr}
 \section{Eight-vertex model}}
 \setcounter{equation}{0}

This was a new  way of solving the six-vertex model. I knew from 
 conversations  with Elliott Lieb that an 
  interesting generalization of the six-vertex model was the 
 eight-vertex model, where one allows two more vertices, one 
 with all arrows in, the other with all arrows out, with weights
 $\om_7, \om_8$, respectively.\footnote{Again, there is a special 
 ``free-fermion'' case, when
 $  \om_1 \om_2 +
      \om_3 \om_4 - \om_5 \om_6  -\om_7 \om_8  =  0$,
  which can be solved by pfaffians.{\color{\ctnclr}\cite{FanWu}}}

 This general eight-vertex model  does not have the ``conservation of down 
 arrows''  property that was necessary for the Bethe ansatz, but an 
 obvious question was whether it has the commuting transfer matrix 
 property. If so, then could one extend the above $T, Q$ 
 relation {\color{\ctnclr}  (\ref{TQ})} to include this model, and so
  calculate   $\kappa$ and the free energy?
  
  It turned out that the answer to both questions was yes, provided
  we restrict consideration to the zero-field model, with
  \be \om_1 = \om_2 = a \sep  \om_3 = \om_4 = b \sep
   \om_5 = \om_6 = c \sep  \om_7 = \om_8 = d \period \ee
   (The last two equalities are not restrictions, as 
   $\om_7$ and $\om_8$ must occur in pairs, as do
    $\om_5$ and $\om_6$.) If we define
 \be \Delta = \frac{a^2+b^2-c^2-d^2}{2(ab+cd)} \sep
  \Gamma = \frac{ab-cd}{ab+cd} \comma \ee
  then the transfer matrices of two models with different
 values of $a,b,c,d$, but the same values of $\Delta$ and 
 $\Gamma$,  commute.
 
 Sutherland had shown in 
 1970{\color{\ctnclr} \cite{Sutherland1970}} 
 that each transfer matrix commutes
 with the hamiltonian of the XYZ chain (with coefficients 
 $\Delta, \Gamma$), which implies the required commutation 
 relation between transfer matrices.
 I was unaware of this, but verified it myself  and went on to 
 construct the matrix $Q(u)$. This also satisfies
  {\color{\ctnclr}  (\ref{TQ})}, but now 
 $\phi(u)$ is a product of Jacobi elliptic theta functions, as  must be
  the eigenvalues $T(u)$, $Q(u)$. They are doubly periodic in the 
  complex $u$-plane  and entire;  $\phi (u)$ and $T(u)$ have 
  $L$ zeros per period  parallelogram, while   $Q(u)$ has $L/2$.  
  Again,   {\color{\ctnclr} (\ref{TQ})} is sufficient to 
  determine the eigenvalues  $T(u), Q(u)$.  The results for the
  critical exponents were unusual and excited interest - they vary 
  continuously with a parameter $\mu$ of the model that 
  corresponds to the $\mu$ of the six-vertex model. For example, 
  the critical exponent $\alpha$ of the specific heat is
  \bd \alpha = 2 - \frac{\pi}{\mu} \period \ed

  The eight-vertex model includes the Ising and six-vertex 
  models as special cases. The Ising case is $\mu = \pi/2$,  
  $\Delta = 0$, corresponding
  to the logarithmic singularity; $\mu = 0$, $\Delta = -1$ 
  corresponds to  the F  model transition, with  $\alpha = - \infty$; 
  and $\mu = \pi$, $\Delta = 1$   to the KDP model transition, with
  $\alpha = 1$.

  Of course, all this took  many weeks to work out, and some of it was 
done  on the ship from England to Australia. I shall always wonder if
I would have had this idea if I had not had that five months' break in 
England -  it can be a good idea to give the mind time to relax and 
go  on auto-pilot.

As with  the Ising model, it was harder matter to obtain the 
spontaneous magnetization and spontaneous polarization.
 I  obtained the polarization in 1973 for the six-vertex model
in the anti-ferroelectric regime $\Delta < - 1$ by a direct 
calculation of scalar products of Bethe eigenvectors:  the method
 was somewhat tortuous, as is indicated by the fact that the
 proof of one step
 depended on  taking $n$ to be a prime number!
 Michael Barber and I conjectured the spontaneous
 magnetization of the eight-vertex model in that 
 year,{\color{\ctnclr} \cite{BarberBaxter1973}} and  
 Stewart Kelland   and I the spontaneous polarization in 
 1974,{\color{\ctnclr} \cite{BaxterKelland1974}}
 but derivations had  to wait for the 
 development of the corner transfer matrix method  in 
 1976.{\color{\ctnclr} \cite{Baxter1976}}



  
{{\color{\ctnclr} \section{Yang-Baxter relation}}
 \setcounter{equation}{0}
 Two-dimensional nearest-neighbour lattice models can be grouped in 
 three classes:
 \begin{itemize}
 \item  Spins live on sites and interact along edges (e.g the 
 Ising model) .
 \item Spins live on edges and interact at a vertex (e.g. the 
 six and eight-vertex models).
 \item Spins live on sites and all four spins (on the square
  lattice) round a face interact (the eight-vertex model can also be 
  formulated this way).
 \end{itemize}
 
  The condition for two transfer matrices to commute is a 
  local one, involving  the Boltzmann weights  $W$ of particular 
  edges, vertices or faces.   It takes different 
  forms{\color{\ctnclr} \cite{BPY1988},\cite[eqns. 9.68, 13.3.1]{book}}
  for the three above  cases, and these can be represented 
graphically  as in Figure {\color{\ctnclr}  \ref{YBreln}}. 
In each case $W_1$ can be 
interpreted as the weight (edge, vertex or face) of one transfer 
matrix, $W_2$ of the 
other, and $W_3$ as a supplementary weight in the equation.
Each figure represents the partition function of a small graph. 
The outer spins (on the open circle sites or the six exterior 
edges of the second figure)
are fixed and the inner ones (solid circles or the three edges of the 
inner triangles)   are to be summed.
The partition functions on each side must be equal, for all
values of the corresponding exterior spins.
  

  The first form is the star-triangle relation, which was used by 
  Onsager.{\color{\ctnclr} \cite{Onsager1944}}  The second was used 
  by    McGuire.{\color{\ctnclr} \cite{McGuire1964}}
  

 \begin{figure}[hbt]
\begin{picture}(320,440) (-35,-290)


\setlength{\unitlength}{0.6pt}

\thicklines


{\color{blue}

\put (65,90) {\large $W_1$}

\put (135,105) {\large $W_2$}

\put (114,36) {\large $W_3$}

\put (51,80) {\line(1,0) {66}}
\put (117,80) {\line(1,3) {20}}
\put (117,80) {\line(1,-1) {40}}

\put (51,79) {\line(1,0) {66}}
\put (118,80) {\line(1,3) {20}}
\put (116,80) {\line(1,-1) {40}}

\put (117,80) {\circle*{12}}
\put (45,80) {\circle{12}}
\put (139,146) {\circle{12}}
\put (162,36) {\circle{12}}

}

{\color{red} 
\put (255,90) {\line(1,0) {20}}
\put (255,89) {\line(1,0) {20}}

\put (255,97) {\line(1,0) {20}}
\put (255,96) {\line(1,0) {20}}
}

{\color{blue}

\put (367,135) {\large $W_3$}
\put (456,95) {\large $W_1$}

\put (390,41) {\large $W_2$}

\put (345,90) {\circle{12}}
\put (441,155) {\circle{12}}
\put (462,47) {\circle{12}}

\put (350,94) {\line(3,2) {85}}
\put (350,86) {\line(3,-1) {106}}
\put (442,148) {\line(1,-5) {19}}

\put (350,93) {\line(3,2) {85}}
\put (350,87) {\line(3,-1) {106}}
\put (442,147) {\line(1,-5) {19}}

}


{\color{blue}

\put (45,-200) {\line(1,2) {85}}
\put (20,-180) {\line(2,1) {180}}
\put (50,-40) {\line(3,-2) {150}}

\put (46,-200) {\line(1,2) {85}}
\put (20,-181) {\line(2,1) {180}}
\put (50,-41) {\line(3,-2) {150}}

\put (69,-175) {\large $W_1$}
\put (115,-79) {\large $W_2$}
\put (144,-140) {\large $W_3$}

}

{\color{red} 
\put (245,-110) {\line(1,0) {20}}
\put (245,-109) {\line(1,0) {20}}

\put (245,-117) {\line(1,0) {20}}
\put (245,-116) {\line(1,0) {20}}
}

{\color{blue}
\put (386,-186) {\line(1,2) {85}}
\put (320,-125) {\line(2,1) {180}}
\put (315,-88) {\line(3,-2) {140}}

\put (387,-186) {\line(1,2) {85}}
\put (320,-124) {\line(2,1) {180}}
\put (315,-87) {\line(3,-2) {140}}

\put (453,-79) {\large $W_1$}

\put (415,-150) {\large $W_2$}

\put (335,-140) {\large $W_3$}

}


{\color{blue}
\put (50,-287) {\line(2,1) {59}}
\put (114,-309) {\line(2,1) {59}}
\put (128,-381) {\line(2,1) {58}}
\put (51,-291) {\line(3,-1) {53}}
\put (121,-256) {\line(3,-1) {53}}
\put (46,-296) {\line(1,-6) {10}}
\put (64,-363) {\line(3,-1) {53}}
\put (110,-318) {\line(1,-6) {10}}
\put (180,-283) {\line(1,-6) {10}}

\put (50,-286) {\line(2,1) {59}}
\put (114,-310) {\line(2,1) {59}}
\put (128,-382) {\line(2,1) {58}}
\put (51,-292) {\line(3,-1) {53}}
\put (121,-257) {\line(3,-1) {53}}
\put (47,-296) {\line(1,-6) {10}}
\put (64,-364) {\line(3,-1) {52}}
\put (111,-318) {\line(1,-6) {10}}
\put (179,-283) {\line(1,-6) {10}}

\put (45,-290) {\circle{12}}
\put (109,-311) {\circle*{12}}
\put (122,-384) {\circle{12}}
\put (115,-255) {\circle{12}}
\put (58,-362) {\circle{12}}
\put (179,-277) {\circle{12}}
\put (191,-349) {\circle{12}}

\put (72,-342) {\large $W_1$}
\put (104,-289) {\large $W_2$}
\put (140,-339) {\large $W_3$}

}

{\color{red} 
\put (235,-320) {\line(1,0) {20}}
\put (235,-321) {\line(1,0) {20}}

\put (235,-313) {\line(1,0) {20}}
\put (235,-314) {\line(1,0) {20}}
}

{\color{blue}
\put (330,-287) {\line(2,1) {59}}
\put (343,-358) {\line(2,1) {59}}
\put (407,-381) {\line(2,1) {58}}
\put (410,-328) {\line(3,-1) {55}}
\put (401,-256) {\line(3,-1) {53}}
\put (326,-296) {\line(1,-6) {10}}
\put (344,-363) {\line(3,-1) {53}}
\put (395,-262) {\line(1,-6) {10}}
\put (460,-283) {\line(1,-6) {10}}

\put (330,-288) {\line(2,1) {59}}
\put (343,-359) {\line(2,1) {59}}
\put (408,-382) {\line(2,1) {58}}
\put (410,-329) {\line(3,-1) {55}}
\put (401,-257) {\line(3,-1) {52}}
\put (327,-296) {\line(1,-6) {10}}
\put (343,-364) {\line(3,-1) {53}}
\put (396,-262) {\line(1,-6) {10}}
\put (461,-283) {\line(1,-6) {10}}

\put (325,-290) {\circle{12}}
\put (406,-327) {\circle*{12}}
\put (402,-384) {\circle{12}}
\put (395,-255) {\circle{12}}
\put (338,-362) {\circle{12}}
\put (459,-277) {\circle{12}}
\put (471,-349) {\circle{12}}

\put (422,-306) {\large $W_1$}
\put (396,-361) {\large $W_2$}
\put (356,-314) {\large $W_3$}

}
 \end{picture}
 \vspace{-1.0cm}
 \caption{\small The three forms of the Yang-Baxter relation.} 
  \label{YBreln}
\end{figure}
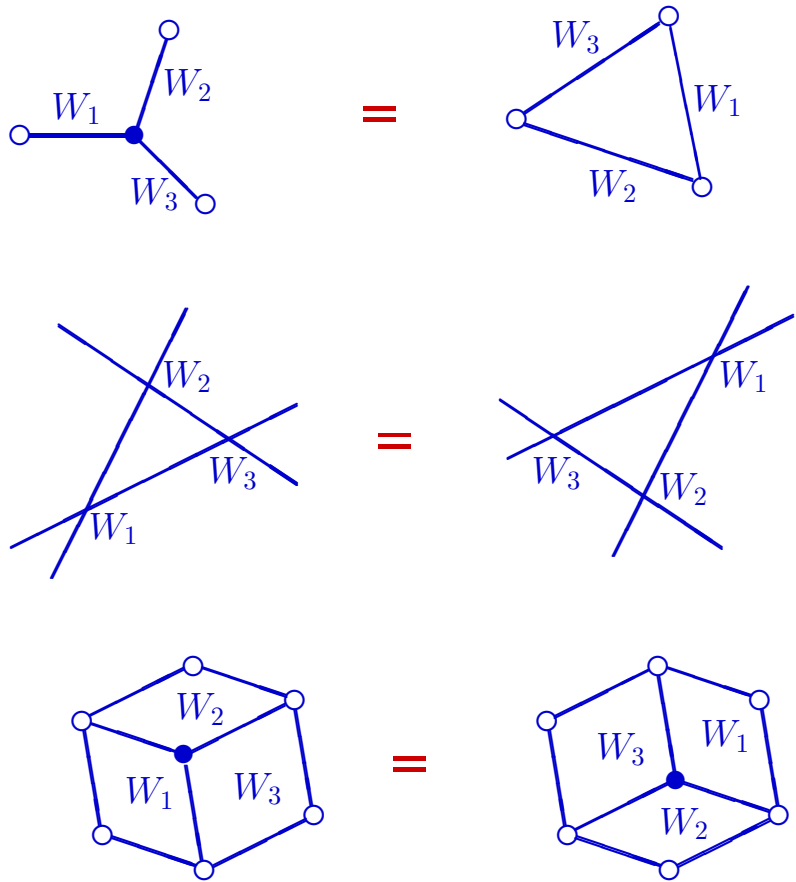


{{\color{\ctnclr}
 \subsection{Other models} }

This general technique turned out to be very useful in identifying
and solving statistical mechanical models: some two-dimensional 
examples are 
\begin{itemize}
\item{Hard-hexagon model in 1980.{\color{\ctnclr} \cite{Baxter1980}}}
\item{Fateev-Zamodchikov model in 
  1982.{\color{\ctnclr} \cite{FZ1982}}}
\item{Kashiwara-Miwa model in 
   1986.{\color{\ctnclr} \cite{KM1986}}}
\item{Solvable $N$-state chiral Potts model in 
  1988.{\color{\ctnclr} \cite{BPY1988}}}
\end{itemize}

In addition, the Yang-Baxter equations can be straightforwardly 
extended to three dimensions. The first solution to the resulting
 ``tetrahedron''  equations was given by Zamolodchikov in 
 1981.{\color{\ctnclr} \cite{Zam1981}}  Further solutions were found
later, e.g by Bazhanov and Baxter.{\color{\ctnclr} \cite{BazBax1993}} 
As a general 
rule these three  dimensional models are critical and their weights
are not necessarily real and positive. This contrasts with
(say) the hard hexagon model, which is a good model of a 
triangular lattice gas, exhibiting the different phases, and with
critical exponents that can (because of universality) be compared
directly with experiment.{\color{\ctnclr} \cite{Schick1982}}

   
{{\color{\ctnclr}
\subsection{Mathematical difficulties}  }
  
  Of all the models mentioned, the Ising model is undoubtedly the 
  simplest. 
 Because of its underlying free-fermion structure, all the eigenvalues
 can be evaluated explicitly, even for a finite lattice width $L$. This 
 contrasts with the other models, where one usually does not know 
 any  explicit solutions of  the non-linear $T, Q$ (or equivalent) 
 relation, and the best one can do is to use analytic techniques to 
 calculate the largest and  near-largest eigenvalues of $T$  in the
  limit  of  large $L$. In general there is no simple direct product 
  property  for the eigenvalues.
 
 
{{\color{\ctnclr}
\vspace{2mm}
 \subsection{Rapidities} }
 
 For all these two-dimensional solvable models, the Boltzmann
  weights  $W$ (and $ \Wb$) depend on both the spin variables 
  (e.g. $a,b $) and on other     ``rapidity'' variables $p, q$:
 \bd  W = W(a,b) = W_{pq}(a,b)  \sep 
  \Wb = \Wb (a,b) = \Wb _{pq}(a,b)  \period \ed 
 These rapidity variables are associated with the dotted lines
  in  Figure {\color{\ctnclr} \ref{sqlatt}}. (For the moment 
  ignore the break in the 
  line below  $\sigma_1$.)   Each edge of the original lattice of
  solid lines   is intersected by two dotted  lines, one horizontal and
   one vertical. The vertical (horizontal) dotted lines have 
 rapidities $p$ or $p'$  ($q$ or $q'$). In general each line may 
 have its
 own rapidity variable, but for an homogeneous model all the
 vertical  rapidities $p$ must be the same,  and similarly for the 
 horizontal rapidities $q$.
 
 The rapidity variables can be chosen so that two row-to-row 
 transfer matrices, with different values $q, q'$ of $q$, commute:
 \bd T_q  \, T_{q'} \eq T_{q'}  \,  T_q \period \ed
If the weight  $W_1$ in Figure {\color{\ctnclr} \ref{YBreln}}
 depends on  two 
rapidities   $q$ and $r $,  and $W_2$ on $r$ and $p$, then 
$W_3$ depends only on $p$ and $ q$:  $W_3$  is independent 
of $r$.


The effect of the Yang-Baxter relation is that one can move these 
rapidity lines around without changing the partition   
function.{\color{\ctnclr} \cite{Baxter1978}} If these moves  do not cross
 the spin $\sigma_1$  in  Figure {\color{\ctnclr} \ref{sqlatt}},  then 
 the RHS of   {\color{\ctnclr} (\ref{defIsingmag})} is  also unchanged.
If the lattice is infinitely big and $\sigma_1$ deep within it,
then any particular rapidity line $p$ (or $q$) can be moved infinitely 
far from $\sigma_1$, where we do not expect $p$  (or $q$)  to 
contribute 
to the RHS of  {\color{\ctnclr}  (\ref{Isingmag})}. It follows that the
 magnetization
$\cal M$ must be independent of {\em all} the rapidity variables, and 
indeed that is implied by the result  
{\color{\ctnclr} (\ref{Isingmag})}. (For the
 Ising model, $K$ and $\overline{K}$ depend on the rapidities,
 but $k$ is a constant, the same for all edges of the lattice.)


 {{\color{\ctnclr}
 \subsection{Rapidity difference property}}

With the exception of the solvable chiral Potts model, all the 
two-dimensional models mentioned above have the {\em rapidity 
difference property}, i.e. their weights $W$, $\Wb$ depend on the
corresponding rapidity variables $p, q$ only via the difference
$p - q$ (the variable $u$ in {\color{\ctnclr} (\ref{TQ})} is such
 a difference). This 
is significant. It means that  $W, \Wb$ are trigonometric or elliptic 
functions of $p-q$. In the limit of a large system the partition 
function per site $\kappa$ can be obtained easily by the ``inversion
relation''  trick{\color{\ctnclr} \cite{Baxter1982}} and the 
spontaneous magnetization by the corner transfer matrix 
method.{\color{\ctnclr} \cite[chapter 13]{book}}
These methods do involve assumptions about the analytic 
properties of the variables, for instance that $\kappa$ is an
analytic function of $u = p-q$ in some vertical strip in the 
complex  $u$-plane.
 
{{\color{\ctnclr}

\vspace{3mm}

\section{Solvable Chiral Potts model}}
 \setcounter{equation}{0}
 
 For the solvable chiral Potts model,  $W(a,b) = W(a-b)$ and 
 $\Wb (a,b) = \Wb (a-b)$, where
 \be W(n) \eq \left( \frac{\mu_p}{\mu_q} \right)^n \, 
 \prod_{j=1}^n \frac{y_q-\omega^j x_p}
 {y_p - \omega^j x_q}  \sep
 \Wb (n) \eq (\mu_p \mu_q )^n \, \prod_{j=1}^n 
 \frac{\omega x_p-\omega^j x_q}
 {y_q - \omega^j y_p}  \period \ee
 

 Here 
  \be \omega \eq \ex^{2 \pi \i /N} \comma  \ee
  $\mu_p, x_p, y_p$ are three complex variables related by
 the two equations
 \be \label{mxy}  k x_p^N = 1 - k'/\mu_p^N \sep 
 k y_p^N = 1 - k' \mu_p^N \comma \ee
and $k, k'$ are constants satisfying
 \be k^2 + {k'}^2 = 1 \period \ee

 We can think of $ p = \{ \mu_p,  x_p,  y_p \}$ as a point on 
 a three-dimensonal
 curve. This is the rapidity $p$. Similarly for $q$.
  If $x_p, x_q, y_p,  y_q$   lie on the unit circle and are 
  ordered anti-clockwise in that sequence, then 
 we can choose $\mu_p, \mu_q$ so that  the Boltzmann weights
 $W, \Wb$ are real and positive. The conditions 
 {\color{\ctnclr}  (\ref{mxy})} for  $p$, $q$ ensure that
 \be
 W(n+N) = W(n) \sep \Wb (n+N) =  \Wb (n)   \period \ee
 The model is therefore $Z_N$-symmetric. It is {\it chiral}
 because $W(-n) \neq W(n)$ and $\Wb (-n) \neq  \Wb (n)$, 
 so reflection-symmetry (left-hand equals right hand) 
 is broken.
 
 
 When $N=2$ this is the Ising model, with $e^{-2K} = W(1)$,
 $e^{-2\overline{K}} = \Wb (1)$ (normalising  the weights  in 
 {\color{\ctnclr} (\ref{Isingwts})}  so that $W(0) = \Wb (0) = 1$).

   For $N > 2 $ this  model does {\em not} have the ``rapidity 
 difference'' property. This makes it much harder to calculate
 the thermodynamic properties. Even so, differential  equations were 
 written down{\color{\ctnclr} \cite{Baxter1988}} in 1988 that in 
 principle
 define $\kappa$ and the critical exponent $\alpha$ was thereby
 found to be $1 - 2/N$.  An explicit expression 
 as a double integral was obtained in 
 1990.{\color{\ctnclr} \cite{Baxter1990,Baxter1991,Baxter2000}}
 

   It is natural to define the magnetization  as 
    \be \label{CPmag}
  {\cal M}_r  \eq  \langle \omega^{ r \sigma_1} \rangle \eq 
  Z^{-1}  \sum_{\sigma} \omega^{r \sigma_1} \, 
   \prod_{\langle  i,j \rangle} W(\sigma_i, \sigma_j)   \ee
for $r = 1, \ldots , N-1$.

  In 1989 Albertini {\it et al } conjectured{\color{\ctnclr} \cite{Alb1989}}
  that for $ 0 < k < 1$
  \be \label{CPconj}
  {\cal M}_r \eq (1-k^2) ^{r (N-r)/2 N^2 } \comma \ee
  which is a beautifully simple formula that fits the $N=2$ Ising
  case and the available series expansions. For $k >1$ the 
  system is disordered and the magnetization vanishes:
  $ {\cal M}_r = 0$.
  
  It was very difficult to prove this conjecture. In his collected
   works C.N. Yang   
  says his 1952 calculation of the  Ising model magnetization  
  took six months and was the longest in his 
  career.{\color{\ctnclr} \cite{Yang1983}}    The chiral Potts 
  magnetization  took much longer: it was not till 2005, after spending 
  much time on  and off mulling  over the problem, that I 
  succeeded{\color{\ctnclr} \cite{Baxter2005}}
  in deriving  the formula   {\color{\ctnclr} (\ref{CPconj})}.
  

{{\color{\ctnclr}
\subsection{Broken rapidity line derivation of ${\cal M}_r$}}
  
  The technique I used was invented by Jimbo, Miwa and 
  Nakayashiki.{\color{\ctnclr} \cite{JMN1993}}  One breaks one of the 
  dotted rapidity 
  lines adjacent to the central spin $\sigma_1$ and gives the two 
  halves different values   $q_1$ and $q_2$ of the rapidity 
  variable,   as in   Figure {\color{\ctnclr} \ref{sqlatt}}.
   
   The effect of this is that one cannot remove these two half-lines
away from $\sigma_1$. One can still remove all the other rapidity
lines,  so now, using the definition
{\color{\ctnclr} (\ref{CPmag})}, \be {\cal M}_r \eq \; {\rm function \; only \; of \; }
k, q_1 \; {\rm and} \; q_2 \period \ee
This is a generalization of the magnetization.

   
   However, one can rotate the two half-lines  and cross them 
 over. This gives
 functional relations for the generalized  ${\cal M}_r$. For the models 
 with the  rapidity difference property,   ${\cal M}_r$ can 
 only depend on  $ q_1, q_2$ via their difference. 
 If one makes plausible assumptions
 about the analyticity properties of this function (e.g. analytic
 in  a  particular vertical strip), then one can solve the functional
 relations  and obtain ${\cal M}_r$. This provides an alternative
 to the corner transfer matrix method of calculating single-spin
 correlations.
 
 Again, however, life is more difficult for the chiral Potts model.
 It does not have the difference property and (ignoring the 
 dependence on $k$, which we regard as a constant)
 ${\cal M}_r$
 is a function of two variables, not one. It is not obvious how to solve
 the functional relations, and what additional information is
 required.


 What I observed in 2005 was that if 
 one took
 \be x_{q_2} \eq x_{q_1} \sep  y_{q_2} \eq \omega  y_{q_1} 
 \sep \mu_{q_2} \eq \mu_{q_1} \comma \ee
and again made a plausible analyticity assumption,
 then one could evaluate  ${\cal M }_r$ for this case. 
 
 Of course one wants it rather for the case when $q_2 = q_1$, but 
 since  it  is then a constant, independent of $q_1$, it is sufficient to 
 obtain it at the intersection of these two cases, when
 \be  y_{q_2 } \eq y_{q_1} \eq 0 \period \ee
  In this way I was able to verify the conjecture 
  {\color{\ctnclr} (\ref{CPconj})}.
  I still do not know what the generalized ${\cal M}_r$ is for arbitrary
  $q_1, q_2$.
  
{{\color{\ctnclr}
 \section{Superintegrable Chiral Potts model}}
 \setcounter{equation}{0}
 We return to the usual situation, where there are no broken 
 rapidity lines  and  the magnetization ${\cal M}_r$ of
  {\color{\ctnclr} (\ref{CPmag})}  is independent of the rapidities.

There is a special  ``superintegrable'' case of the chiral Potts
model, when  the vertical rapidities alternate, as in
 Figure{\color{\ctnclr}  \ref{sqlatt}},
taking the values  $p, p', p, p', \ldots $, where
\be x_{p'} \eq y_p \sep y_{p'} \eq x_p \sep \mu_{p'} \eq 1/\mu_p 
\period \ee
Since $\cal M$ is independent of the rapidities, its value
for the superintegrable case is the same as that of 
 the general solvable chiral Potts model.
 
  The chiral Potts model may be the most difficult of the 
 two-dimensional  solvable models, but its  superintegrable 
 case has some remarkable
 simplifications. In fact if we impose cylindrical boundary conditions 
 as in Figure {\color{\ctnclr} \ref{sqlatt}}, with the top and bottom spins fixed to zero,
 then it has properties similar to those of the 
 Ising model.
 (For $N=2$, the superintegrable case, like the general solvable
 case, {\em is} the general zero-field  Ising model.)

 
 With these boundary conditions, the partition function $Z$ is 
 given by {\color{\ctnclr} (\ref{partfn2})} and ${\cal M}_r$ by
 \be \label{formM}
  {\cal M}_r \eq \frac{ u^{\dagger} T^n S_r T^{\, n'} u }{Z} 
 \comma \ee
 where $n$ is the number of rows below $\sigma_1$ and 
 $n'$ the number above, so $n + n' = M$. The vector 
 $u$ is defined by {\color{\ctnclr} (\ref{defu})}.   Also, 
 $S_r$ is a diagonal matrix
 with entries
 \be \left(S_r \right)_{s,s'}  \eq \omega^{r \sigma_1 } 
 \prod_{i=1}^L \delta(\sigma_i, \sigma'_i) \ee
 again writing the spin set 
$\{ \sigma_1, \ldots \sigma_L \}$ as $s$.  

{{\color{\ctnclr}
\subsection{Partition function $Z$}}


Define a set of vectors $u_0, \ldots, u_{N-1}$ with elements
\be \left( u_a  \right)_s \eq \prod_{i=1}^L \delta(\sigma_i, a)
 \period \ee
Then $u = u_0$, where $u$ is the vector above, defined by 
{\color{\ctnclr} (\ref{defu})}.

 Let $R$ be the spin-shift matrix with elements
\be R_{s,s'} \eq \prod_{i=1}^L \delta( \sigma_i, \sigma'_i +1 ) 
\period \ee
Then  $R^N = 1$ and the eigenvalues of $R$ are $1, \omega,
\ldots, \omega^{N-1}$. Let ${\cal V}_Q$ be the vector space
of all vectors $v$ such that $Rv = \omega^Q v $. Then 
$u_a = R^a u $ and 
\be \label{defvQ}
v_Q \eq  N^{-1/2} \, \sum_{a=0}^{N-1} \omega^{- Qa} u_a
\; \;  \; \in \; {\cal V}_Q   \ee
 and
  \be Z = Z_0 + Z_1 + \cdots + Z_{N-1} \comma  \ee
where
 \be Z_Q \eq Z_Q(M) \eq  v_Q^{\dagger} T^M v_Q  \period \ee

 These matrices are of dimensions $N^L$ and ${\cal V}_Q$ is of 
 dimension $N^{L-1}$. However, if we repeatedly multiply
 $v_Q$ by $T$, we generate a smaller 
 sub-space,{\color{\ctnclr} \cite{RJB1988, RJB1989}} of dimension
 $2^m$, where
 \be m \eq m_Q \eq \left[ \frac{(N-1)L - Q}{N} \right] 
\ee
and we  write $[x]$ for the integer part of $x$.
Further, we can choose the basis vectors (independently of $q$ 
and  $k$) so that
\be T \eq t_1 \otimes t_2 \otimes \cdots \otimes t_m \ee
and 
\be  v_Q \eq 
\left( \begin{array}{c} 1 \\  0 \end{array} \right)  
 \otimes \cdots \otimes 
 \left( \begin{array}{c} 1 \\  0 \end{array} \right)  \period \ee
 Here each $t_j$ is a two-by-two matrix. So to calculate
 $Z_Q$ we only need $T$ in  a $2^m$-dimensional sub-space,
 within which it is a  direct product, as in the Ising 
 model. 
 
It follows   that $Z_Q =  Z_Q(M) $ is a product of $m$ 
 simple factors.
 

{{\color{\ctnclr}
\subsection{Magnetization}}
From {\color{\ctnclr} (\ref{formM})} and 
{\color{\ctnclr} (\ref{defvQ})},
\be {\cal M}_r \eq \frac{ W_{0,r} + \cdots + W_{N-1,r}}
{Z_0 + \cdots+ Z_{N-1}} \comma \ee
where
\be W_{Q,r} \eq v_Q^{\dagger} T^{\, n} \, S_r T^{\, n'} \, 
v_{Q+r} \period \ee


When $N=2$ we regain the Ising model and we know 
then{\color{\ctnclr} \cite{Yang1952, MPW1963,paper1}} for finite $L$ 
that
$W_{Q,r}$ can be written as a determinant. We have just seen that
the superintegrable  chiral Potts  resembles the Ising model
in that $Z_Q$ is a simple product. We therefore ask whether 
our $W_{Q,r}$ can be written as a determinant for all $N$?

The answer is yes. It can be written as a determinant.  The working is 
given in a series of papers (not necessarily  in logical 
order).{\color{\ctnclr} \cite{paper2, paper3,paper4,paper5,Iorgov2010}}

Setting $m' = m_{Q+r}$, we find that 
\be  \label{resdet}
W_{Q,r} \eq Z_Q(n) Z_{Q+r}(n') \det [ 1 + A A^T ] \comma \ee
where $A$ is an $m$ by $m' $ matrix.

This is huge progress. If $n= 3$ and $L = 15$, then $m = m' = 9$. We 
have reduced the problem from one calculating the elements of powers
of the transfer matrix $T$, of dimension 14,349,907, to one of 
evaluating a nine by nine determinant!

Even so,  we do want to take the large lattice limit. 
It is easy to allow $n, n'$ to become infinite - we merely take the 
limit of the elements of the matrix $A$. We then want to 
let $L \rightarrow \infty$.

In this limit even $m$ and $m'$ will become infinite, so we need a 
way of evaluating the determinant in  {\color{\ctnclr} (\ref{resdet})}.

This was a tricky problem and I mulled over it for more than a year.
If  $m= m'$, then $A$ is square and invertible, so 
\be \det [1 + A A^T]   \eq \det A \; \det [{( A^T)}^{-1} + A] \period \ee
The elements of $A$ are of the form
\be \label{Cauchy}
A_{ij} \eq \frac {x_i x'_j}{c_i - c'_j} \comma \ee
where $x_i,  x'_j, c_i, c'_j $ are known parameters.  

The matrix $A$ is therefore Cauchy-like
and its determinant is{\color{\ctnclr} \cite{Krattenthaler2001}}
\be  \label{detCauchy}
\det A \eq  \frac { X X'  \prod_{1 \leq i  <  j \leq m }(c_i - c_j)  (c_j-c'_i)}
{ \prod_{i=1}^m \prod_{j=1}^m (c_i-c'_j) } \comma \ee
where $X = \prod_i x_i$,  $X'  = \prod_j x'_j $.
Further, the inverse of the transpose of a Cauchy-like matrix is also 
Cauchy-like,  the elements also being of the form
{\color{\ctnclr}  (\ref{Cauchy})}, with only $x_i, x'_j$ changed to some 
values $y_i, y'_j$.


The elements of the desired sum are therefore 
\be  \label{expr}
\left( ({A^T)}^{-1} + A \right)_{ij} \eq 
\frac {y_i y'_j +x_i x'_j } {c_i - c'_j} \period \ee

This is {\em not} a Cauchy-like matrix, rather it is akin to a 
Pick matrix of displacement rank 2.{\color{\ctnclr} \cite{Agler2002}}  
There are fast computational
algorithms for numerically calculating  the determinants of such 
matrices,{\color{\ctnclr} \cite[Chapter 1]{Kailath1999}} but I know 
of no explicit  expression for the answer.


However, I finally looked at the numerator in the desired limit
$n, n' \rightarrow \infty$. It is a rational function of  
$\lambda_i, \lambda'_j $, where 
\be c_i = (1+k'^2-{\lambda_i}^2)/2 k'  \sep
 c'_j =  (1+k'^2-{\lambda'_j}^2)/2 k'  \ee
 so
  \be c_i - c'_j \eq ({\lambda'_j}^2 - \lambda_i^2 )/2 k'
  \period  \ee
Further, the numerator of {\color{\ctnclr} (\ref{expr})} turns 
out to be of the form
  \be y_i y'_j + x_i x'_j \eq s_i s'_j (\lambda_i +\lambda'_j ) \ee
  so the factor $\lambda_i +\lambda'_j $ {\em cancels out} 
   of   {\color{\ctnclr} (\ref{expr})}, leaving
  \be  \label{expr2}
\left( ({A^T)}^{-1} + A \right)_{ij} \eq 
\frac {- 2 k' s_i s'_j  } {\lambda _i - \lambda'_j} \period \ee
 This {\em is} a Cauchy-like matrix and one can  obtain 
 its determinant from the general formula 
 {\color{\ctnclr}  (\ref{detCauchy})}.
 In this way one can obtain $W_{Q,r}/ Z_Q(n)Z_{Q+r}(n')$
 as a double  product over $i,j = 1, \ldots m$, for
 {\em finite} $L$.
 
 In general $m, m'$ differ by at most one. If they are different,
 one can add a  row or column to $A$ so as to make it square
 and still Cauchy-like, while leaving $1 + A A^T$ unchanged. 
 The same cancellation then  occurs in 
 {\color{\ctnclr}  (\ref{expr})} and again 
 one obtains the double product expression for $W_{Q,r}$.
 
 Finally one lets $m, m' \rightarrow \infty$ and evaluates the double
 product, using  a Wiener-Hopf factorization. We again of course
 obtain the formula {\color{\ctnclr}  (\ref{CPconj})}.
 
So this provides a rigorous proof of the magnetization of the 
superintegrable chiral Potts model, and hence of the Ising 
model. It does not involve Szeg{\H o}'s theorem. The  full
calculation is given in refs.{\color{\ctnclr} \cite{paper1} -- \cite{paper5} }
and {\color{\ctnclr} \cite{Iorgov2010}}.
A related derivation of the magnetization via the pair correlation 
function is given in {\color{\ctnclr} \cite{Perk2010}}.

{{\color{\ctnclr} 
 \section{Summary}}
 \setcounter{equation}{0}
 

There have been many developments in the statistical mechanics of 
lattice models since Onsager's famous solution of the Ising model in 
1944. The general trend has been one of increasing complexity, first
to models  without the direct product transfer matrix property, then to 
the chiral Potts model where one loses also the useful 
rapidity-difference property.

  
But if one uses cylindrical boundary conditions with fixed spins on 
the top and bottom rows, then the  wheel comes full circle with the 
superintegrable case of the chiral Potts model. One obtains 
determinantal expressions for the magnetization, analogous to those 
found for the Ising model by 
Onsager{\color{\ctnclr} \cite{KaufmanOnsager1949}},
Yang{\color{\ctnclr} \cite{Yang1952}}  and Montroll, Potts and 
Ward{\color{\ctnclr} \cite{MPW1963}}.

One can evaluate these 
determinants, not by calculating matrix eigenvalues,
nor by using  Szeg{\H o}'s theorem, but by identifying
the determinants as Cauchy-like.

 
{{\color{\ctnclr}

\vspace{3mm}

 \section{Acknowledgements} }
 \setcounter{equation}{0}

 The author thanks  Jacques Perk for telling him of 
 ref.{$\,$\color{\ctnclr}\cite{JSP1995}} and  the material on the
the Lars Onsager Online archive  at \\
{\color{\ctnclr} {\footnotesize 
http://www.ntnu.no/ub/spesialsamlingene/tekark/tek5/arkiv5.php}}.\\ 
He also thanks Harold Widom for sending him a copy of the letter 
 {\color{\ctnclr} \cite{JSP1995}}, Richard Askey for alerting him to
 page {\color{\ctnclr} 41} of Onsager's collected works, and Henk van Beijeren
 for helpful comments on the six-vertex model.

{{\color{\ctnclr}

 \end{document}